\documentclass[fleqn,usenatbib]{mnras}

\usepackage{newtxtext,newtxmath}

\usepackage[T1]{fontenc}
\usepackage{ae,aecompl}

\usepackage{lscape}

\usepackage{graphicx}	
\usepackage{amsmath}	
\usepackage{amssymb}	

\def\Swift{{\em Swift}}
\newcommand\XMM{{\em XMM-Newton}}
\newcommand\ergscm{$\rm erg\,cm^{-2}\,s^{-1}$}

\newcommand\source{SWIFT\,J0746.3-1608}

\title[The nature of Swift J0746.3-1608]{The true nature of Swift J0746.3-1608: a possible Intermediate Polar showing accretion state changes}

\author[F. Bernardini et al.]
{F.~Bernardini,$^{1,2,3}$\thanks{E-mail:federico.bernardini@inaf.it} 
D.~de Martino,$^{2}$ 
K.~Mukai,$^{4,5}$
M.~Falanga,$^{6,7}$
\\
$^1$ INAF - Osservatorio Astronomico di Roma, via Frascati 33, I-00040 Monteporzio Catone, Roma, Italy \\
$^2$ INAF $-$ Osservatorio Astronomico di Capodimonte, Salita Moiariello 16, I-80131 Napoli, Italy\\
$^3$ New York University Abu Dhabi, Saadiyat Island, Abu Dhabi, 129188, United Arab Emirates\\
$^4$ CRESST and X-Ray Astrophysics Laboratory, NASA Goddard Space Flight Center, Greenbelt, MD 20771, USA\\
$^5$ Department of Physics, University of Maryland, Baltimore County, 1000 Hilltop Circle, Baltimore, MD 21250, USA\\
$^6$ International Space Science Institute (ISSI), Hallerstrasse 6, 3012 Bern, Switzerland\\
$^7$ International Space Science Institute Beijing, No.1 Nanertiao, Zhongguancun, Haidian District, 100190 Beijing, China}

\date{Accepted XXX. Received YYY; in original form ZZZ}

\pubyear{2018}

\begin{document}
\label{firstpage}
\pagerange{\pageref{firstpage}--\pageref{lastpage}}
\maketitle

\begin{abstract}

Optical and X-ray observations suggested that the 9.38 h binary, \source\, could be a Cataclysmic Variable of the magnetic or nova-like type, or a low mass X-ray binary. Its optical, UV, and X-ray light curves are strongly variable over years. We report on a recent \XMM\ observation (28 April 2018), when the source had recovered from a deep low state that likely begun mid-late 2011. We detect for the first time a signal at about 38 min that we interpret as the rotation of the accreting white dwarf primary. Its amplitude decreases with increasing energy, indicating localised photoelectric absorption from cold material. The X-ray spectrum shows optically thin thermal emission with excess at the iron complex, absorbed by a dense medium partially covering the X-ray source. Based on these features, we propose that \source\ is a magnetic CV of the Intermediate Polar (IP) type. The long-term light curves at different wavelengths show high and low states, a rare phenomenon in the IP subclass and observed so far in only three other systems. The long orbital period, the peculiar long term variability, and its proposed magnetic nature, makes \source\ an interesting evolutionary test case.

\end{abstract}




\section{Introduction}

Intermediate Polars (IPs) are a subclass of Magnetic Cataclysmic Variables (MCVs), compact binaries where a magnetic (B$\leq10^{7}$ G) white dwarf (WD) primary accretes matter from a Roche lobe overflowing main sequence or sub-giant secondary \citep{ferrario15,mukai17}. Accretion onto the WD can occur through a disc or directly from a stream, depending on the magnetic field intensity and degree of asynchronism. IPs are usually desynchronized systems for which ${\rm P_{spin=\omega}<P_{orb=\Omega}}$. Furthermore, a hybrid  accretion in the form of disc overflow \citep{hellier95,norton97}, is also possible, and frequently observed \citep[see e.g.][]{bernardini12,bernardini17}. In these accretion models, matter close to the WD surface is channeled along the magnetic field lines, and reaching supersonic velocities is forming a post-shock region (PSR) above the surface. The PSR is hot ($kT\sim10-80$ keV) and in IPs the flow cools and slows down mainly via bremsstrahlung radiation (hard X-ray) \citep{aizu73,wu94,cropper99}. This X-ray emission is then pulsed at the spin period of the WD (disc-accretion), at the beat (disc-less) or at both periods in the case of disc overflow \citep[][] {hellier95}. 
The vast majority of IPs are persistent systems, but a small subgroup including FO Aqr, AO Psc, and V1223 Sgr, showed low accretion states \citep[][]{garnavich88,kennedy17b,littlefield18}. Such fading is believed to be due to a temporarily reduction of the mass transfer rate from the donor star \citep{livio94}.

\source\ (hereafter J0746) is one of the still unclassified sources in the high-energy catalogs \citep{oh18,bird16}. Optical follow-ups classified J0746 to be either a MCV or a nova-like system with a long, 9.38 h, orbital period. In X-ray, the "Neil Geherls" \Swift\ satellite (hereafter \Swift) observed the source to be highly variable \citep{thorstensen13,parisi14}. 
When first observed by \XMM\ in 2016, it was at its lowest recorded flux level preventing a classification. Its X-ray spectral characteristics suggested either a MCV or a LMXB \citep[][]{bernardini17}. An X-ray monitoring with \Swift\ allowed us to detect a flux increase starting from November 2017, indicating the source had recovered from its lowest state and was rising towards the higher state. We here report on detailed timing and spectral analysis of a second \XMM\ observation performed on 28 April 2018, when the source was a factor of about 20 brighter than in 2016. We complement the study with optical/UV and X-ray long-term light curves from the \Swift\ UVOT and XRT instruments. These show that J0746 is a peculiar CV, likely a MCV of the IP type, which shows rapid and particularly intense accretion states changes.

\section{Observation and data reduction}

J0746 was observed on 28 April 2018 at 15:44:21 UTC by \XMM\ (obsid 0830190701) with 
the European Photo Imaging Cameras \citep[EPIC: PN, MOS1 and MOS2][]{struder01,turner01,denherder01} complemented with simultaneous optical monitor \citep[OM,][]{mason01} photometry in the U-band. The PN camera operated in timing mode, while the two MOS cameras in small window mode (thin filter always applied). The PN exposure was shorter than those with the MOS cameras (20 vs 22 ks). Data were reduced and inspected for all instruments, but due to the source nature (e.g. slow rotator), we only report on the MOS Cameras. Data were processed using the Science Analysis Software (\textsc{SAS}) version 16.1.0 and the latest calibration files available in 2018 May. Source photon event lists and spectra were extracted from a circular region of radius 40 arcsec. The background was extracted from a region free from sources contamination in the same CCD where the source lies. High particle background epochs were removed for the spectral analysis while, for the timing analysis, the entire dataset was used. RGS spectra from the standard SAS pipeline were used. The OM operated in fast window mode using the
U-band (2600--4300 \AA) filter (21.4 ks exposure). Light curves were extracted using the standard pipeline and corrected to the Solar System barycenter using the \textsc{barycen} task.

The task \textsc{epiclccorr} was used to generate background-subtracted light curves in the 0.3--2, 2--3, 3--5, 5--12, and the whole 0.3--12 keV band.  Spectra were rebinned using \textsc{specgroup} imposing a minimum of 30 counts in each bin and a maximum oversampling of the energy resolution by a factor of three. Spectra were also extracted at the minimum and maximum of the spin cycle. MOSs spectra were fitted simultaneously by using \textsc{Xspec} version 12.9.1p package \citep{arnaud96}. XRT (0.3--10 keV) and UVOT (U and UW1 band) \Swift\ light curve and spectra were also extracted. For the XRT data we used the products generator available at Leicester \Swift\ Science Centre \citep{Evans09}, for the UVOT data we followed standard procedures (http://www.swift.ac.uk/analysis/uvot/).

\section{Data analysis and results}

\subsection{Timing analysis} 
\label{sec:timing}
\begin{figure}
\begin{center}
\includegraphics[angle=-90,width=9cm]{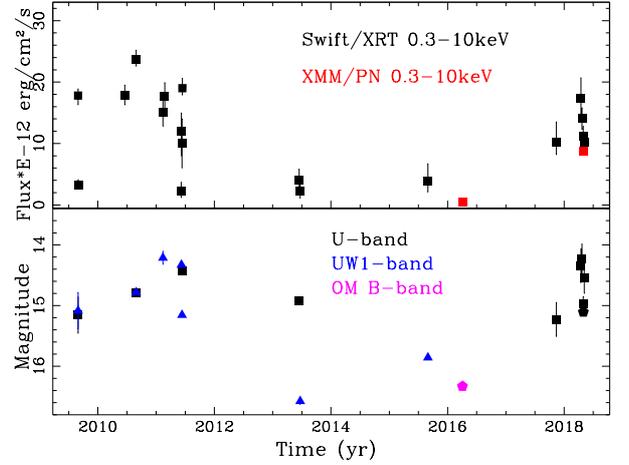} 
\caption{Long-term optical, UV, and X-ray light curves of \source\ from \Swift. \XMM\ average pointing flux is also plotted in red.}
\label{fig:lc-long}
\end{center}
\end{figure}

We first inspected the long-term optical, UV, and X-ray light curves recorded by \Swift\ (Figure \ref{fig:lc-long}) binned with one point per obsid (typically there are a few snapshots about 500--1000 s long in each obsid). 
J0746 is highly variable both on short (less than a day) and long timescales (more than a day). Low and high flux states are clearly present, where we arbitrarily define ${\rm F_{Low}<5\times10^{-12}}$ and ${\rm F_{High}>5\times10^{-12}}$ \ergscm. The 0.3--10 keV flux was derived using the automatic pipeline by extracting one spectrum per obsid.
Strong conclusions about state transitions (and duration) can not be made due to the sparse coverage. However, transitions seem to be rapid. On August 29, 2009 the flux decreased within less than a day from high ($1.7\times10^{-11}$ \ergscm) to low ($\sim3.3\times10^{-12}$ \ergscm) state (first two points on the left of Figure \ref{fig:lc-long}). The opposite, fast, transition (low to high) was recorded in June 2011. J0746 was then found in the low state both in 2013 and 2015, and when observed by \XMM\ in April 2016, where it was at its lowest recorded level ($\sim5.6\times10^{-13}$ \ergscm). The \Swift\ monitoring that started in November 2017 showed the source rising towards higher levels which was confirmed in additional pointings in April-early May 2018. The second \XMM\ observation took place in between the last two \Swift\ observations, confirming the recovery to a high state.
\noindent Flaring-like behaviour with amplitude $\sim10$ was recorded during the high state, where the 0.3--10 keV flux can vary from $\sim9\times10^{-12}$ to $\sim9\times10^{-11}$ \ergscm, for example \citep[see also Figure 7 in][]{thorstensen13}. 
The UV and optical light curves are broadly correlated to the X-ray light curve, with the UVW1-band ($\sim 2200-4000$ \AA) flux that was more than two magnitudes dimmer in 2013 with respect to 2011 (16.6 vs 14 mag). The XMM-Newton pointings showed the source at B$\sim16.3$ mag in 2016 and at U$\sim15.1$ mag in April 2018. Although observed in different energy bands the OM data confirms the source brightening in 2018.

\begin{figure}
\begin{center}
\includegraphics[angle=-90,width=9cm]{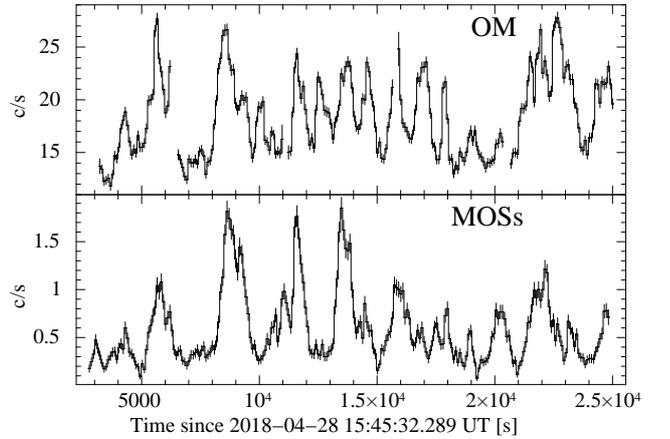} 
\caption{28 April 2018 OM U-band and MOSs 0.3--12 keV light curves binned at 90 s.}
\label{fig:lc-xmm}
\end{center}
\end{figure}

Both the EPIC MOSs event data and background subtracted light curves were summed together to study the time variability. The 0.3--12 keV and U-band light curves (Figure \ref{fig:lc-xmm}) are both strongly variable, with multiple peaks repeating on short timescale, where the 0.3--12 keV flux averagely changes by a factor of about 10 (2 in the U-band). Assuming a constant spectral shape (Sect. \ref{sec:spec} and Table \ref{tab:spec}) the flux changes from a minimum of $5\times10^{-12}$ \ergscm (0.1 c/s) to a maximum of $9\times10^{-11}$ \ergscm (2 c/s). These strong changes are similar to those observed by XRT during the high state in August 2009. This indicates that J0746 is a strongly variable source at all energies, even during the high states.

\begin{figure}
\begin{center}
\includegraphics[angle=270,width=3.2in]{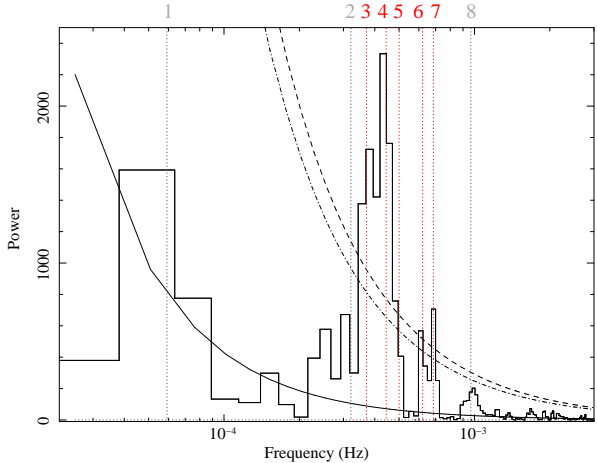} 
\caption{The power spectrum of the combined MOS photon events of J0746 in the soft 0.3-2 keV band, where the spin signal is strongest. Sideband frequencies are calculated using $P_{\omega}=2253$ s (this work) and $P_{\Omega}=9.3841$ h \citep{thorstensen13}. The left solid line is the fit to the log-log periodiogram with bias removed, the dot-dahsed and dashed lines mark the 95 and 99 per cent global confidence 
levels, respectively. The numbers from 1 to 8 correspond to the frequencies used to fit the light curve. Statistically significant frequencies are in red, while frequencies below 99 confidence level are in grey. See Section \ref{sec:timing} for more details.}
\label{fig:powspec}
\end{center}
\end{figure}

While no clear sign of a long-term periodic variation linked to the orbital period (9.38 h) is present, the strong variability over thousands of seconds hints at periodic signals.  
The power spectra were then computed in different energy ranges at the temporal resolution of the MOS cameras (0.3 s). They all show a structured peak in the range 0.35--0.5 mHz and a few weaker ones at both higher and lower frequencies.  The lowest-frequency peak (P1, see Figure \ref{fig:powspec}), although compatible with half of the 9.38 h orbital period, is as discussed below, not statistically significant and likely due to red-noise. The first harmonic of the orbital frequency (2$\Omega$) was found to dominate the B-band light curve during the low state in 2016, but not in the X-rays \citep{bernardini17}. The broadness of the main peak hints at the presence of multiple contributions of close periodicities. Therefore, we attempted to determine them using two different methods, concentrating on the softest 0.3--2 keV band.
Firstly, we used the {\sc ftools} task 
{\sc efsearch}\footnote{https://heasarc.gsfc.nasa.gov/ftools/}. A Gaussian fit gives peak centroids at: P$_{2}=3258\pm80$, P$_{3}=2700\pm100$, P$_{4}=2253\pm40$, P$_{6}=1611\pm8$, P$_{7}=1459\pm40$, and P$_{8}=1034\pm24$ s \footnote{With this method, P$_{5}$ can not be separated from P$_{4}$.}. All uncertainty are hereafter reported at the $1\sigma$ confidence level. 
We also fitted the 0.3--2 keV light curve binned at 10 s resolution using the {\sc Period04} package \citep{lenz05} that allows to perform multi-sinusoidal fits by using the relevant frequencies identified by detrending the light curve iteratively. A fit with eight frequencies gave a reasonable reproduction of the light curve, although not satisfactory  ($\chi^2_{\nu}=1.53$, dof=2080).  
The uncertainties in the parameters were evaluated using Markov chain Monte Carlo (MCMC) simulations within {\sc Period04} package, performing 500 iterations. While P$_{1}$ and P$_{6}$ result unconstrained, we found: P$_{2}=3202\pm79$ s, P$_{3}=2736\pm58$, P$_{4}=2311\pm45$, P$_{5}=2010\pm37$, P$_{7}=1466\pm8$, and P$_{8}=1102\pm23$. The latter values are  consistent within uncertainties with those measured with {\sc efsearch}. 
The significance of the peaks in the power spectrum, which is affected by red noise, was also evaluated. We followed \cite{vaughan05} assuming a power-law for the underlying spectrum (P$_{\nu}=N\nu^{-\alpha}$).  The log-log periodiogram with bias removed was then fitted with a linear function with slope $\alpha= 2.10\pm0.18$ ($\chi^{2}_{\nu}=1.47$, dof=367). We then evaluated the 95 and 99 per cent global confidence levels using 369 independent frequencies. Only the two broad peaks around 0.4 and 0.7 mHz (encompassing P$_{3}$ to P$_{7}$) are detected above the 99 per cent level (Figure \ref{fig:powspec}).

\begin{figure}
\begin{center}
\includegraphics[angle=270,width=3.0in]{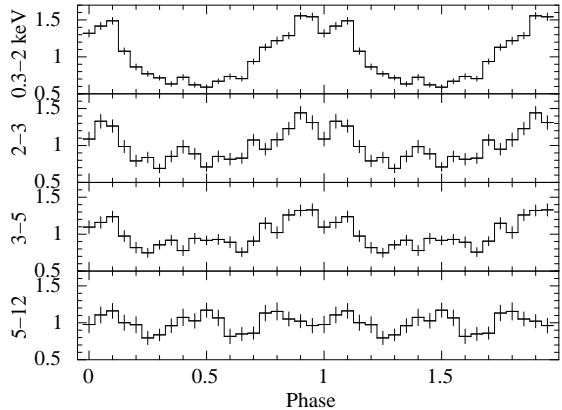} 
\caption{Background subtracted light curve in different energy bands folded at the spin period (2253 s). The reference folding time is the integer of the observation starting time. On the y-axis the Normalized Intensity is plotted. Two cycle are shown for plotting purposes.}
\label{fig:puls_vs_e}
\end{center}
\end{figure}

The pulsed fraction (PF) \footnote{$\rm PF=(F_{max}-F_{min})/(F_{max}+F_{min})$, where $\rm F_{max}$ and $\rm F_{min}$ are the maximum and minimum fluxes of the sinusoid at the fundamental frequency, respectively.} was also computed as a function of the energy interval. We found that all signals have a constant PF with the exception of P$_{4}$, where it significantly decreases with increasing energy (PF$_{0.3-2}=43.2\pm1.2 $, PF$_2-3=24.3\pm2.4$, PF$_{3-5}=18.9\pm2.2$, and PF$_{5-12}=3.0\pm3.0$ per cent; Figure \ref{fig:puls_vs_e}). A typical characteristics of MCVs, and of IPs in particular, is the presence of a energy dependent spin periodicity, ascribed to photoelectric absorption from cold material within the magnetically confined accretion flow \citep{rosen88}.
Assuming P$_{4}$ as the spin ($\omega$) of the accreting WD, combining it with the orbital period from \cite{thorstensen13}, we find for the significant periodicities that 1/P$_{3}\sim\omega-2\Omega$, 1/P$_{4}\sim \omega$, 1/P$_{5}\sim\omega+2\Omega$ and 1/P$_{7}\sim2(\omega -3\Omega)$, while no interpretation is found for P$_{6}$. To our knowledge, P$_{7}$ is not known to be strong in any other IPs.
The spin is the highest signal in the 0.3--2 keV band and emerges from the broad peak in the region that would encompass $\omega-\Omega$, $\omega+\Omega$, and $\omega+2\Omega$ (Figure \ref{fig:powspec}). We emphasize that a longer exposure would have improved the resolution allowing to separate $\omega$ from its closer sidebands. In conclusion, we propose that P$_{4}\sim2300$ s (about 38 min) is the spin period of the WD. We note that during the low state an indication of an X-ray periodicity at 2700 s was found, whose origin was unclear at that time \citep{bernardini17}, but we find it now consistent with $\omega-2\Omega$.

In the U-band power spectrum, the strongest signal is P$_{3}\sim2800$ s, and P$_{6}\sim1575$ and P$_{7}\sim1450$ s are also present, while there is no power at P$_{4}$ ($\omega$).  

To inspect spectral variation as a function of the spin period, the hardness ratio (defined as the count rate ratio in each phase bins between two selected energy ranges) were computed. The comparison between 0.3--2 and 3--5 keV clearly shows that the spin modulation is harder at minimum (HR$\sim0.5$) while it is softer (HR$\sim0.3$) at maximum.

\subsection{Spectral analysis}
\label{sec:spec}

J0746 spectrum is thermal and optically thin, with an excess at the iron complex (6.4--7.0 keV), as is typically observed in MCVs. These sources are characterized by multi-temperature spectra absorbed by dense cold material localized within the binary system, likely in the accretion column above the shock \citep[see e.g.][]{Done95,EzukaIshida99,bernardini13,mukai15,bernardini17,bernardini18}.
The 0.3--10 keV spectrum was then fitted with a model made of 
an optically thin plasma component (\textsc{mekal} 
or \textsc{cemekl} in \textsc{Xspec}), where metal abundances (A$_{\rm Z}$) with respect to Solar, set to that of the ISM from \cite{wilms00}, are left free to 
vary, plus a narrow Gaussian line fixed at 6.4 keV to account for the fluorescent Fe K$_{\alpha}$ line, all absorbed by a total (\textsc{phabs}) and a partial (\textsc{pcfabs}) covering columns. The use of \textsc{pcfabs} is justified by the fact that the PF decreases as the energy increases. 
We did not fit simultaneously the BAT spectrum accumulated over many years since it would require a cross normalization component of about 3. This implies that J0746 has been on average brighter than during the 2018 \XMM\ pointing.

Both the multi-temperature plasma (\textsc{cemekl}) and the single temperature plasma (\textsc{mekal}) models
provide an acceptable fit to the data ($\chi^{2}_{\nu}=1.05$, 192 d.o.f., $\chi^{2}_{\nu}=1.06$ 193 d.o.f., respectively; Table \ref{tab:spec} and Figure \ref{fig:spec}). The total absorber is two order of magnitude lower than that of the ISM in the source's direction \citep{kalberla05}, suggesting a nearby source. The partial covering absorber has a column density ${\rm N_{H_{Pcf}}}=8-9\times10^{22}$ cm$^{-2}$ and a covering fraction cvf$\sim45$ per cent. A soft blackbody component (kT$_{\rm BB}\sim20-100$ eV), frequently found in the spectrum of MCVs, is not statistically required. The maximum plasma temperature of \textsc{cemekl} is 
kT$=18\pm^{5}_{2}$ keV, higher but consistent within $2\sigma$ with that measured with \textsc{mekal}. We stress that the former should be considered as a more reliable lower limit to the shock temperature \footnote{The \textsc{cemekl} maximum temperature slightly increases to $20.7\pm4.0$ keV when including the BAT spectrum and allowing for the large cross normalization.}. 
We also derive a lower limit to the WD mass by using a more physical model developed by \cite{suleimanov05}, which takes into account temperature, density, and gravity gradients within the PSR. We included in the fit the BAT spectrum as the model relies on the high energies to derive the mass. The eight-channel spectrum from the first 70 months of BAT monitoring \citep{baumgartner13} available at the NASA GSFC site was used. The spectral analysis was restricted to 80 keV. A cross normalization constant and a broad Gaussian were added to account for flux variability and the iron line complex, respectively. The fit gives M$_{\rm WD}=\rm0.78\pm_{0.13}^{0.08}\, M_{\odot}$ ($\chi^{2}_{\nu}=1.13$, 195 d.o.f.). 

\begin{figure}
\begin{center}
\includegraphics[angle=0,width=8cm]{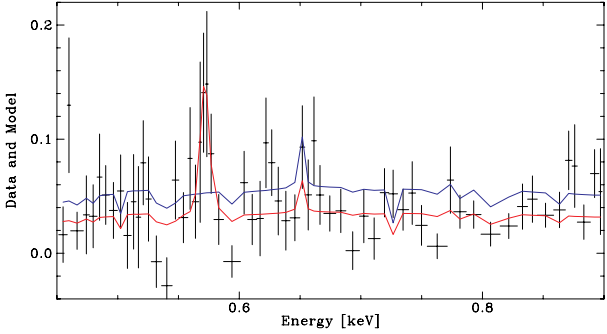} 
\caption{RGS1 data in the 0.45--0.9 keV range, grouped to have a minimum of 9 source counts per bin, are plotted against energy.
RGS2 data are not shown since the spectral range around OVII line (0.57 keV) falls in a gap due to a dead CCD chip. 
The blue line is the best-fit \textsc{cemekl} model from the EPIC data analysis, with cross normalization constant fixed to 1.
The red line is the model with the cross normalization constant left free to vary, and with a Gaussian with EW$=$24 eV added.}
\label{fig:rgs}
\end{center}
\end{figure}

\begin{figure*}
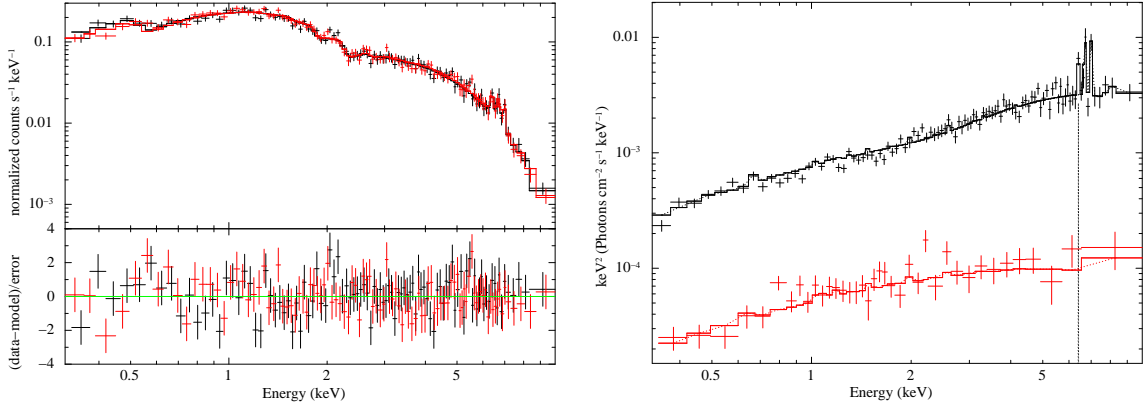

\begin{center}
\begin{tabular}{cc}
\includegraphics[angle=270,width=3.0in]{avspec_mek.eps} 
\includegraphics[angle=270,width=3.0in]{high_vs_low_mek.eps} 
\end{tabular}
\caption{Left: The folded 0.3--10 keV average spectrum using \textsc{mekal} (MOS1 black, MOS2 red), residuals are shown in the lower panel. Right: Comparison between the MOS1 unfolded average spectra of the high-state (black) and low-state (red). The low-state spectrum is slightly softer.}
\label{fig:spec}
\end{center}
\end{figure*}

The RGS data show the presence of OVII line centered at $0.572\pm0.003$ keV, whereas the best-fit \textsc{cemekl} model hardly predicts any flux from this line (Figure \ref{fig:rgs}). The equivalent width is 24 (8--40) eV, where the error is for 90\% confidence level, fully accounting for the uncertainties in both the Gaussian and \textsc{cemekl} components. Similar excess of OVII line has been seen in several other IPs \citep[e.g. HT Cam,][]{demartino05}. The centroid energy could suggest a combination of resonant and intercombination components, while the forbidden component appears very weak. The detection  OVII line clearly indicates that the post-shock region is multi-temperature, supporting our preference for the \textsc{cemekl} model.

To investigate the role of spectral parameters in generating the X-ray spin modulation, a (spin-)phase resolved spectroscopic analysis was performed. Spectra extracted at spin minimum ($\phi=0.3-0.6$) and maximum ($\phi=0.8-1.1$) were fitted separately using both models in Table \ref{tab:spec}. N$_{\rm H_{Ph}}$, A$_{\rm Z}$, were fixed at their average spectrum values. All other parameters were left free to vary. For both models we found that the spin modulation is due to an increase of the covering fraction at spin minimum (from $35\pm4$ to $64\pm5$ per cent) and of the normalization at spin maximum (by $\sim20$ per cent). As a last step, we compared the average high state spectral results with that from the low state (Figure \ref{fig:spec}, right panel). We used the data presented in \cite{bernardini17} and re-fitted them with both \textsc{mekal} and \textsc{cemekl}, fixing N$_{\rm H_{Ph}}$ and A$_{\rm Z}$ to the average, high state, best fit values. We did not include the \textsc{pcfabs} and \textsc{Gaussian} components in the latter fit, because neither is statistically required. During the low state the source is softer, with kT$_{\rm mekal}=7\pm1$ keV, or kT$_{\rm cemekl}=10.9\pm1.5$ keV (F$_{\rm X,bol}\sim5.6\times10^{-12}$ \ergscm).

\begin{table*}
\caption{Best fit models to the average 0.3--10 keV spectrum. 
The absorbed 0.3--10 keV and unabsorbed bolometric 
(0.01--200 keV) fluxes are also reported. 
Uncertainties are at $1\sigma$ confidence level.}
{\small
\begin{center}
\tabcolsep=0.18cm
\begin{tabular}{ccccccccccc}
\hline 
mod. & N$_{\rm H_{Ph}}$  & N$_{\rm H_{Pcf}}$  &  cvf & kT  & n           & A$_{\rm Z}$ & EW  & F$_{0.3-10}$ & F$_{\rm X,bol}$ &  $\chi^2$/dof  \\
     & $10^{22}$         & $10^{22}$          &      &     &  $10^{-3}$  &             &     & $10^{-12}$   & $10^{-12}$      &   \\
     & cm$^{-2}$         & cm$^{-2}$          & \%   & keV &             &             & keV & erg/cm$^2$/s & erg/cm$^2$/s    &   \\         
\hline 
cemek$^a$ & 0.013$\pm$0.005 & 8.2$\pm$12  & 45$\pm$3 & $18\pm_{4}^{5}$ $^b$       & 23$\pm_{4}^{7}$  & 0.78$\pm$0.17 & 0.11$\pm$0.02 & 8.7$\pm$0.2 & $\sim18.4$ & 1.05/192 \\
mek       & 0.015$\pm$0.005 & 9.3$\pm$1.2 & 46$\pm$3 & $11.3\pm_{1.1}^{1.5}$ & 6.2$\pm$0.3      & 0.67$\pm$0.14 & 0.11$\pm$0.01 & 8.7$\pm$0.1 & $\sim17.3$ & 1.06/193 \\
\hline
\end{tabular}  
\label{tab:spec}                      
\end{center}} 
\begin{flushleft}
$^a$ Multi-temperature power-law index $\alpha=2\pm_{0.4}^{0.9}$.\\
$^b$ Maximum temperature.\\
\end{flushleft}                               
\end{table*}

\section{Discussion and conclusions}

The April 2018 \XMM\ observation has shown that J0746 had recovered from its low state since mid-late 2011. Its orbital period is firmly established to be 9.38 h from optical spectroscopy \citep{thorstensen13}.
We found an extremely variable X-ray source and identified a prominent periodicity at $\sim2300$ s, which we interpret as the spin period of the accreting WD. Due to the short X-ray coverage, we are unable to separate this periodicity from the beat, $\omega-\Omega$, and $\omega+\Omega$, the sidebands most commonly found in IPs power spectra. 
If it is the spin period of a magnetic WD, then J0746 is an IP with a spin to orbital period ratio {\rm P$_{\omega}/P_{\Omega}\sim 0.07$}, fully consistent with the observed ratios in the majority of IPs \citep{norton04,norton08,bernardini17}.
It shows a strongly variable X-ray light curve. 
The spin modulation is the strongest signal in the soft X-rays, as typically found in IPs due to the effects of photoelectric absorption in the magnetically confined accretion curtain. We also identified close-by sidebands, likely due to $\omega-2\Omega$, $(\omega+2\Omega)$, and $2(\omega-3\Omega)$, with possibly others combining together. Although not statistically significant at low frequencies, the signal at $2\Omega$ appears to affect the power spectrum also at higher frequencies. It was found to also dominate the optical light curve during the low state \citep{bernardini17}.
The X-ray power spectra of IPs are often complex, with multiple sidebands between the spin and orbital frequency and combination of harmonics \citep{norton89}.
Theoretical power spectra reproducing the multiplicity of spin, sidebands, and harmonics \citep{norton96} as those observed in J0746 would require a complex geometry, likely involving non-symmetric opposite poles unequally contributing. This makes \source\ power spectrum particularly interesting for future works to solve the accretion geometry in a likely more complex magnetic field configuration.

We derived the distance of J0746 from the ESA GAIA DR2 data release \citep{gaia16,gaia18a}.
J0746 has a well constrained trigonometric parallax $\pi=1.54\pm0.03$ mas which, using space density prior based on the Galactic tridimentional model as seen by Gaia \citep[see details in][]{bailer-Jones18}, translates into a distance $d=638\pm12$ pc.
Using $d$ and the low and high state bolometric fluxes (F$_{\rm X,bol}$), assuming $L_{acc}=GM\dot{M}/R\sim L_{X,bol}$, and using M$_{\rm Wd}=0.78{\rm M_{\odot}}$, we obtain the following mass accretion rates: $\dot{M}_{High}\sim9.8\times10^{-11}\,M_{\odot}\,yr^{-1}$ and 
$\dot{M}_{Low}\sim3.0\times10^{-12}\,M_{\odot}\,yr^{-1}$, a change of factor $\sim33$. 
Therefore, J0746 is a rare X-ray selected MCV showing extreme high and low X-ray states. 
In that respect, J0746 is reminiscent of the IP FO Aqr \citep{littlefield16,kennedy17b}, where $\Delta\dot{M}\sim7$. In addition, \cite{simon15} reports on the optical and Swift/BAT low states of another IP, V1223 Sgr, and a optical low state of AO Psc is also known \citep{garnavich88} although, in this case, we do not know if it was accompanied by an X-ray low state. Finally, the peculiar, long orbital period CV, Swift J1907.3-2050=V1082 Sgr, also exhibits high and low states in the optical and flaring behavior in its X-ray light curve \citep{Thorstensen2010,bernardini13}. If the IP classification is correct, it makes J0746 only the second IP, after FO Aqr, to have been observed in X-rays in high and low states. This kind of state transitions in CVs are thought to be due to changes in the mass transfer rate from the donor star. These are possibly due to surface magnetic spots that concentrates in front of the inner Lagrangian point (L1) partially filling it, so temporarily reducing the mass transfer \citep{livio94}. The presence of star spots has been inferred for the first time in the donor of the IP AE Aqr \citep{hill16}, although it has never been observed to undergo state changes. Star spots have been claimed to be responsible for the low states of Polars and of the state changes in FO Aqr \citep{kennedy17b}. We note that, interestingly, star spots at L1 should produce random (e.g. not periodic) state transitions \citep{hessman00}, as we indeed observed in J0746. We suggest that in the low accretion state the disc is depleted and the accretion occurs mainly by a stream. Moreover, a negligible absorption is expected in the X-rays due to the reduced mass accretion rate onto the WD. Indeed during the low state, the source is faint, showing no sign of intrinsic absorption, and only the sideband ($\omega-2\Omega$) is detected at X-ray wavelengths, while there is no power at the spin, indicating that accretion directly impacts the WD without an intervening disc. At optical wavelengths, $2\Omega$ is detected, suggesting a strong contribution of ellipsoidal modulation of the donor and consequently a faint disc. During the high state instead, likely the disc is well developed, accretion proceeds from the disc along the field lines over the polar regions, and the local absorption is much higher (N$_{\rm H}>10^{22}$ cm$^{-2}$). The presence of several sidebands of the spin and orbital periods would also suggest a contribution of disc-overflow \citep{hellier95}.
Our results on the very recent recovery to the high accretion state in J0746 and its possible IP classification show that the occurrence of high and low states, although rare in this class of MCVs, is a phenomenon common to all types of CVs.  Particularly interesting is J0746 long orbital period. The three confirmed IPs observed so far to display high/low states have all shorter orbital periods, with AO Psc and V1223 Sgr in the 3--4 hr range of VY Scl stars, and FO Aqr at 4.85 h (while the peculiar CV V1082 Sgr has 20.8 h). V1223 Sgr has shown extended ($\sim10$ yr) low states and the other two systems have undergone short epochs of low accretion rates, with FO Aqr slowly recovered from a low state in 2016 in about a hundred of days \citep{garnavich88,littlefield16}. In May 2018, FO Aqr showed its third low state in 2 years, despite it was always found at a  constant luminosity level before 2016 for about 100 years \citep{littlefield18}. J0746 is at 9.38 h and shows the fastest state transitions (less than a day). This makes this source (and long period systems in general) particularly interesting since their donors should be nuclear evolved \citep{goliasch15} and thus could be test cases to understand angular momentum loss governing their evolution towards short orbital periods \citep{spruit83,andronov03}. State transitions remain a complex still not understood phenomenon observed not only in CVs, but also in LMXBs that needs clear explanation. J0746 recently returned to a strongly variable high state, and follow up multi-band monitoring, possibly capturing state transitions, would be particularly important.

\section*{Acknowledgements}

We thank Dr. Norbert Schartel and the \Swift\ team for granting DDT \XMM\ and \Swift\ time, respectively. We acknowledge useful discussion with Gianluca Israel. We thank the anonymous referee for the useful comments. FB is founded by the European Union's Horizon 2020 research and innovation programme under the Marie Sklodowska-Curie grant agreement n. 664931. DdM is supported from the Italian Space Agency and National Institute for Astrophysics, ASI/INAF, under agreements ASI-INAF I/037/12/0 and ASI-INAF n.2017-14-H.0. This work is based on observations obtained with \XMM, an ESA science mission with instruments and contributions directly funded by ESA Member States, with \Swift, a NASA science mission with Italian participation, and with \textit{Gaia}, a ESA mission. \textit{Gaia} data are processed by the Data Processing and Analysis Consortium (DPAC). 

\bibliographystyle{mnras}
\bibliography{biblio} 

\bsp	
\label{lastpage}
\end{document}